\documentstyle[11pt,aaspp4]{article}
\input psfig

\begin{document}
\title{Gamma-Ray Burst Afterglow: Polarization and Analytic Light Curves}
\author{Andrei Gruzinov \& Eli Waxman}
\affil{Institute for Advanced Study, School of Natural 
Sciences,
Princeton, NJ 08540}

\begin{abstract}

GRB afterglow polarization is discussed. We find an observable, 
up to $\sim 10\%$, polarization, if the magnetic field coherence length 
grows at about the speed of light after the field is generated at the 
shock front. Detection of a polarized afterglow would show that 
collisionless ultrarelativistic shocks can generate strong large scale 
magnetic fields and confirm the synchrotron afterglow model. Non-detection, 
at $\sim 1\%$ level, would imply that either the synchrotron 
emission model is incorrect, or that strong magnetic fields, after they 
are generated in the shock, somehow manage to stay un-dissipated 
at ``microscopic'', skin depth, scales. Analytic lightcurves of synchrotron  
emission from an ultrarelativistic self-similar blast wave are obtained for 
an arbitrary electron distribution function, taking into account the effects 
of synchrotron cooling. The peak synchrotron flux and the flux at 
frequencies much smaller than the peak frequency are insensitive to the 
details of the electron distribution function; hence their observational 
determination would provide strong constraints on blast wave parameters. 

\end{abstract}

\section{Introduction}

X-ray, optical and radio emission following gamma-ray bursts (GRBs) are 
in broad agreement with models based on relativistic blast waves at 
cosmological distances (\cite{AGWa,AGWMR,AGV,AGR,AGKP,AGSPN}). In these 
models, the energy released by an 
explosion, $\sim10^{52}$erg, is converted to kinetic energy of a thin baryon 
shell expanding at ultra-relativistic speed. After producing the GRB, the 
shell impacts on surrounding gas, driving an ultra-relativistic shock 
into the ambient medium. In what follows, we refer to the surrounding gas 
as interstellar medium (ISM) gas, although the gas need not necessarily be 
inter-stellar. The expanding shock continuously heats fresh gas 
and accelerates relativistic electrons, which produce the observed 
afterglow radiation through synchrotron emission (\cite{PRAG,MRAG,VAG}).

To match the observations, the magnetic field behind the shock has 
to be $\sim 10\%$ of equipartition with the shock-heated, compressed 
ISM. What is the origin of this field? The shock-compressed ISM field is many orders of magnitude 
smaller than needed. The magnetic field frozen into the initial GRB 
fireball loses strength by the time the afterglow stage begins, and it is in a wrong place 
anyway. During the afterglow, the decompressed GRB field is located 
far behind the shock, while most of the energy is in the recently 
shocked ISM. Therefore, the magnetic field most likely must be generated 
in and by the blast wave. If the coherence length of the generated field 
is comparable to the thickness of the blast wave, the radiation will 
be polarized. An $\sim 10\%$ degree of polarization is expected. This is significantly smaller than the maximal synchrotron polarization, $\sim 70\%$, because the emitting region is thin and broad; it must be covered by $\sim 100$ mutually incoherent patches of magnetic field. 

In a paper on microlensing of GRB afterglows, Loeb \& Perna (1998) have mentioned the possibility that the afterglows are polarized. Here we estimate the degree of polarization (\S 4). This paper also provides (\S 3)
exact analytic afterglow lightcurves for an arbitrary electron distribution 
function, including the effects of electron cooling. In \S 2 we describe the underlying model assumptions. We discuss the implications of our results to afterglow observations in \S 5. Most of the details of our derivations are given in appendices A--E.

\section{The blast wave model}

A strong blast wave is fully specified by two parameters: the blast wave 
energy $E=10^{52}E_{52}~{\rm erg}$, and the ISM density $n_i=1n_1~
{\rm cm}^{-3}$. With  sufficient 
accuracy, the unshocked ISM may be taken to be cold unmagnetized hydrogen. To calculate the synchrotron emission we need to know 
the fraction of energy in magnetic fields $\xi _B$, and in electrons 
$\xi _e$, and the shape of the electron distribution function (a function 
$f_e(z)$ with first two moments equal to 1). We include $\xi _B$, 
$\xi _e$, and $f_e(z)$ in the list of independent parameters. In 
principle, these are determined by the blast wave energy and the 
ISM density, but a theory of strong collisionless shocks is not 
available (Sagdeev 1966, Krall 1997). 

The plasma flow in the shocked ISM is assumed to be described by the 
Blandford-McKee (\cite{BnM}) self-similar solution. The Lorentz factor 
of the shock wave $\Gamma$, the Lorentz factor of the flow $\gamma$, 
the proper energy density $e$, and the proper number density $n$ 
for all space-time points in the shocked plasma are given in Appendix A. 

We assume that magnetic fields and electrons are described by 
simple scalings. The magnetic field is $B^2/8\pi=\xi_Be$, and 
$\xi _B$ is the same in all space-time points. The electron distribution 
function in the local rest frame has the same shape in all space-time 
points on the shock front, after the shock passage it evolves by adiabatic and synchrotron cooling. At the shock front, the mean energy of an electron in the 
local rest frame is $\gamma _em_ec^2=\xi _ee/n$, and $\xi _e$ is constant. The shape of the electron distribution 
function is not specified at this point. We include $f_e(z)$ into the 
definition of the synchrotron emission function $F$. The synchrotron 
power per unit frequency per electron emitted in the local rest frame is
\begin{equation}
P(\omega )={\sqrt{3}\over 2\pi}{e^3B\over m_ec^2}F({\omega \over \omega _c}),
\end{equation}
\begin{equation}
\omega _c={3\gamma _e^2eB\over 2m_ec},
\end{equation}
but $F$ {\it is not} the standard dimensionless synchrotron 
emission function given in Rybicki \& Lightman (1979). $F$ depends
 on the shape of the electron distribution function (B7). 

Given the set of blast wave parameters, we will measure time, frequency, 
and spectral luminosity (energy per time per frequency) in units of 
\begin{equation}
T=c^{-1}\left( {17\over 8\pi }{E\over n_im_pc^2}\right) ^{1/3}=5.5\times 
10^7\left( {E_{52}\over n_1}\right) ^{1/3}{\rm s},
\end{equation}
\begin{equation}
\omega _0\equiv 3\sqrt{\pi }\left( {m_p\over m_e}\right) ^{5/2}{c\over r_e}\xi 
_B^{1/2} \xi _e^2 (n_ir_e^3)^{1/2}=3.9\times 10^{10}\left( {\xi _e\over 
0.1}\right) ^2\left( {\xi _B\over 0.1}\right)^{1/2}n_1^{1/2}{\rm s}^{-1},
\end{equation}
\begin{equation}
E_0\equiv {17\over 2\sqrt{6\pi }}\left( {m_e\over m_p}\right) ^{1/2}\xi 
_B^{1/2} (n_ir_e^3)^{1/2}E=2.2\times 10^{31}\left( {\xi _B\over 
0.1}\right)^{1/2}n_1^{1/2}E_{52}{\rm erg}.
\end{equation}
The formal origin of these units is explained in Appendices A, B. Their 
physical meaning is illustrated by the following order of magnitude 
statement. At observed time $t_o/T=1$, the blast wave slows down to 
Lorentz factor 2; it radiates at frequency $\omega/\omega_0=1$,
 with spectral luminosity $L/E_0=1$. Our analysis is restricted to the 
ultrarelativistic stage, that is to dimensionless observed times $t_o\ll 1$.

\section{Lightcurves}

We first calculate in \S3.1 synchrotron emission of the blast wave 
neglecting radiative cooling of electrons, i.e. assuming that 
the shape $f_e(z)$ is the same in the 
entire shocked plasma. In \S3.2 we relax this assumption: $f_e(z)$ is determined at the shock front and evolves by 
synchrotron and adiabatic cooling thereafter.

Our analytic lightcurves are exact under the 
following assumptions: (i) The blast-wave hydrodynamics is described by 
the Blandford-McKee self-similar solution (\cite{BnM}); (ii) The 
magnetic field energy density is a fixed fraction of the total energy 
density, independent of space and time; (iii) The electron distribution 
function is determined at the shock front and evolves afterwards 
only through adiabatic and synchrotron cooling.  \cite{GPS} numerically derived exact 
lightcurves for power-law electron distribution, under the assumptions 
described above and neglecting electron cooling. It should be emphasized 
that since a theory of strong collisionless shocks is not available 
at present, none of the above assumptions can be justified. Thus, the 
numerical values (e.g. of the peak flux and peak frequency as function 
of time) derived here under these assumptions are not necessarily more 
accurate than those obtained by order of magnitude estimates
(e.g. \cite{AGWb,AGSPN,AGWG}).

The exact analytic lightcurves are useful because they allow us to determine which afterglow characteristics are strongly dependent on the details of the electron distribution function, and which are insensitive to these details and depend mainly on the global blast wave parameters (i.e the blast wave energy, the ambient medium density, and the energy fractions carried by electrons and magnetic fields). 

\subsection{Adiabatic Lightcurve}
In Appendix B, we derive an expression for the spectral luminosity, neglecting 
synchrotron cooling of electrons. At observed time $t_o$ after the 
gamma-burst, at frequency $\omega$, distant observer (with negligible 
redshift) infers a selfsimilar narrow-band luminosity of the blast wave 
$L_{\omega}(t_o)=L_A(\omega t_o^{3/2})$, where
\begin{equation}
L_A(\omega )=48\int_0^1 da ~a^3(1+7a^2)^{-2}F[~2a(1+7a^2)\omega ~].
\end{equation}

We show the lightcurves in Fig.1. The three curves correspond to different 
doubly normalized electron distribution functions:
\begin{enumerate}
\item Power law, index $p=2.4$: ~~$f_e=f_P=31.2z^2(1+122z^{p+2})^{-1}$ 

\item Maxwellian: ~~~~~~~~~~~~~~~~~~~~$f_e=f_M=13.5z^2e^{-3z}$

\item Mixed: ~~~~~~~~~~~~~~~~~~~~~~~~~~~~~$f_e=0.7f_M+0.3f_P$

\end{enumerate}
Note, that our ``power-law'' distribution, for which $f_e\propto z^{-p}$ 
for $z\gg1$, includes a low energy tail, $f_e\sim z^2$ for $z\ll1$. We 
believe the inclusion of such a ``thermal'' low energy tail is more 
realistic than assuming a sharp cutoff of the electron distribution 
below a certain minimum $z$ value.

\subsection{Lightcurve with synchrotron cooling}
At early times or at high frequencies, synchrotron cooling of the 
electron distribution function will have a noticeable effect on the 
lightcurve. We calculate the nonadiabatic lightcurve, 
$L_{\omega }(t_o)=L_{NA}(\omega, t_o)$ in Appendix C, 
neglecting effects of cooling on the blast wave propagation:
\begin{equation}
L_{NA}(\omega, t_o)=192\int_0^1dy~y^3\int_0^1 da ~a^3(1+7a^2)^{-2}\int 
dz_0~f_e(z_0)~F_0[~2a(1+7a^2)\omega t_o^{3/2}z^{-2}~],
\end{equation} 
where
\begin{equation}
z^{-1}=z_0^{-1}+A(8t_o)^{-1/2}a^{-1}y^{-2}(1-y^{19/6}),
\end{equation}
and
\begin{equation}
A={8\over 19}\left( {m_p\over m_e}\right) ^{2}\sigma _TcTn_i\xi _B\xi 
_e=1.6\times 10^{-2}\left( {\xi _e\over 0.1}\right) \left( {\xi _B\over 
0.1}\right) E_{52}^{1/3}n_1^{2/3}.
\end{equation}
Scaled spectra at different observed time $t_o$ are shown in Fig. 2 for the power-law (and in Fig. 3 for the mixed) electron distribution function of \S 3.1 and $A=0.01$. At high frequencies, eqs. (7), (8) predict a power-law luminosity $L\propto \omega ^{-p/2}$ for an electron distribution function with a power-law tail of index $p$.

\subsection{Observables}
From Eqs. (3)-(5) and Figs. 2, 3, an afterglow at redshift $z_b$, 
observed $t_{\rm day}$ days after the $\gamma$-burst, will show 
maximal flux at a frequency 
\begin{equation}
\nu _m\sim3\times 10^{12}{\sqrt{1+z_b} \over \sqrt{2} }\left( {\xi _e\over 
0.1}\right) ^2\left( {\xi _B\over 0.1}\right)^{1/2}E_{52}^{1/2}t_{\rm 
day}^{-3/2}{\rm Hz}.
\end{equation}
The maximal flux does not depend on the time of observation. 
In a flat universe with Hubble constant $H_0=75~{\rm km/s/Mps}$, 
\begin{equation}
F_{\nu m}=4\left( \sqrt{1+z_b} -1\over \sqrt{2}-1\right) ^{-2}\left( {\xi 
_B\over 0.1}\right)^{1/2}n_1^{1/2}E_{52}{\rm mJy}.
\end{equation}
As seen in Figs. 1, 2, 3, the peak flux $F_{\nu m}$ is robust, i.e. it is 
independent of the details of the electron distribution function. 
The flux below the peak is also robust, and for $\nu \ll \nu _m$ it is given by
\begin{equation}
F_{\nu}=0.6\left( \sqrt{1+z_b} -1\over \sqrt{2}-1\right) ^{-2}\left( {1+z_b\over 2}\right) ^{-1/6}\left( {\xi _e\over 
0.1}\right) ^{-2/3}\left( {\xi _B\over 0.1}\right)^{1/3}n_1^{1/2}E_{52}^{5/6}t_{\rm 
day}^{1/2}\nu _{\rm GHz}^{1/3}{\rm mJy}.
\end{equation}

The peak frequency $\nu _m$ is model-dependent, and may differ by an 
order of magnitude between different electron distribution functions 
with similar $\xi_e$. The spectral shape (equivalently the time profile) 
above the peak is also strongly dependent on the details of the electron 
distribution function.

\section{Polarization}
Synchrotron radiation is highly polarized (Ginzburg 1989), but for this 
polarization to be measurable in an unresolved source, the magnetic field 
coherence length should be comparable to the source size. Here we show that 
GRB afterglows might be polarized. An $\sim 10\% $ polarization seems to be an 
upper bound, corresponding to a coherence length that grows at about the 
speed of light after the field is generated at the shock front.

Qualitatively, our polarization analysis can be summarized as follows. Suppose 
that the magnetic field coherence length in the local rest frame is $l\sim 
c\tau$, where $\tau$ is the proper time after the shock. The extension of the 
emitting region transverse to the line of site is $\sim 5c\tau$. There are 
$\sim 50$ coherent patches. The  degree of polarization is $\sim 60\% 
/\sqrt{50}\sim 10\% $. If the coherence length is smaller than the proper 
time, $l\sim \epsilon c\tau $, $\epsilon <1$,  the degree of polarization 
is decreased to
\begin{equation}
\Pi\sim 10\epsilon ^{3/2}~\% .
\end{equation}
The degree and direction of polarization should depend on time, the 
polarization coherence time should be $\sim \epsilon t_o$. 

\subsection{Magnetic field generated by a relativistic collisionless shock}
As far as we know, magnetic field generation in collisionless shocks is not 
understood. It seems possible that, at the shock front, Weibel instability 
generates near equipartition ($\xi _B\sim 0.1$) small scale ($\sim$ skin depth 
$\delta$, here $\delta=c/\omega_{p}$, $\omega_{p}^2\sim ne^2/\gamma 
_pm_p\lesssim ne^2/\gamma _em_e$) magnetic fields. By magnetic moment 
conservation,  electrons are accelerated to near equipartition energies 
(~relativistic version of Sagdeev 1966, Kazimura et. al. 1998). The ultimate 
fate of the field many skin depth behind the shock front is not clear. What 
happens to the magnetic field coherence length $l$, and to the magnitude $B$ 
at a distance $\Delta \gg \delta$ behind the shock front? Three scenarios seem 
to make sense:
\begin{enumerate}
\item The generated field dies out after finishing its job of isotropizing the 
plasma and bringing it to a state given by the shock jump conditions.

\item The magnitude stays at about equipartition, the coherence length stays 
at about the skin depth.

\item The magnitude stays at about equipartition, the coherence length grows 
as $l\sim \Delta$.
\end{enumerate}

Scenario 1 is not consistent with the synchrotron emission model for the 
afterglow, because too little synchrotron radiation is produced by a skin-deep 
shell with strong magnetic fields. Scenario 2 means unpolarized radiation. We 
will evaluate the degree of polarization for scenario 3.

\subsection{Coherent patch}
Assume that two events in the shocked ISM belong to the same coherent patch if 
the difference in their proper times elapsed after the shock passage $\delta 
\tau$, and their spatial separation transverse to the line of site $\delta h$ 
are small:
\begin{equation}
\delta \tau<\epsilon _{\tau }\tau ,
\end{equation}
\begin{equation}
\delta h <\epsilon _h c\tau ,
\end{equation}
$\epsilon _{\tau },\epsilon _h<1$, $\tau$ is the averaged proper time since 
the shock. By the proper time of an event in the shocked ISM we mean the 
proper time after the shock of a fluid particle at this event. 

\subsection{Degree of Polarization}
As shown in Appendix D, for the emission event with observed time $t_o$, the 
proper time $\tau$ (in units of $T$) and the transverse distance (in units of 
$cT$) are 
\begin{equation} 
\tau=1.15t_o^{5/8}a^{5/4}y^{1/4}(1-y^{9/4}),
\end{equation}
\begin{equation} 
h=1.83t_o^{5/8}a^{1/4}(1-a^2)^{1/2}y^{1/4}.
\end{equation}
Here the meaning of dimensionless variables $a$, $y$ is unimportant, what 
matters is that the luminosity is given by the integral (B20) over $a$ and $y$:
\begin{equation}
L(\omega )=192\int_0^1dy~y^3\int_0^1 da ~a^3(1+7a^2)^{-2}F[~2a(1+7a^2)\omega 
~].
\end{equation}

Now we can separate the full luminosity (18) into coherent parts according to 
the criterion (14),(15). This gives, approximately, the degree of 
polarization. By numerical simulations (Appendix E),
\begin{equation}
\Pi\sim 10\epsilon _h\epsilon _{\tau }^{1/2}~\% .
\end{equation}

\section{Summary of results}

\subsection{Lightcurves}

We have derived simple,
exact analytic afterglow lightcurves for an arbitrary electron distribution 
function, including the effects of electron synchrotron cooling, Eq. (7), 
and in the limit where synchrotron cooling is negligible, Eq. (6).
Our lightcurves are exact under the 
following assumptions: (i) The blast-wave hydrodynamics is described by 
the Blandford-McKee self-similar solution (\cite{BnM}); (ii) The 
magnetic field energy density is a fixed fraction of the total energy 
density, independent of space and time; (iii) The electron distribution 
function is determined at the shock front and evolves afterwards 
only through adiabatic and synchrotron cooling.

We have shown, see. Fig. 1, 2, 3, that the peak synchrotron flux, Eq. (11), and
the synchrotron flux
at frequencies well below the peak flux, Eq. (12), are
insensitive to the details of the electron distribution function. Since the peak flux is also insensitive
to the details of the blast wave hydrodynamic profiles (Fig. 1 in \cite{AGWc}),
the peak flux and the flux at frequencies well below the
peak depend mainly on the global blast wave parameters:
blast wave energy, ambient medium density, magnetic field and electron
energy fractions [cf. Eqs. (11,12)]. Observational determination of these 
fluxes would therefore provide strong 
constraints on blast wave parameters. The numerical value of the peak flux 
derived here, Eq. (11), is similar to that derived in \cite{GPS}, within 
a factor $\sim3$ of the values given in
\cite{AGWb} and \cite{AGWG}, and a factor $\sim10$ smaller than given in
\cite{AGSPN}. 
The discrepancy with \cite{AGSPN} is mainly due to the fact
that it is assumed by these authors that the spectral width of the observed
spectrum at fixed time is determined mainly by the intrinsic spectral
width of synchrotron emission, while the actual width is significantly larger
and dominated by contribution to the observed spectrum at given time 
from different space-time points with different plasma conditions.

The peak frequency and spectral shape at frequencies above the peak are
strongly dependent on details of the electron distribution function, see
Figs. 1, 2, 3.  Furthermore, the peak frequency is also strongly dependent on the 
details of the blast wave hydrodynamic profiles (Fig. 1 in \cite{AGWc}).
This, and the fact that the spectral peak is flat, imply that 
observational determination of the peak frequency and of spectral
features above the peak at a given time can not be used directly to
constrain global blast wave parameters. These features would mostly provide
information on the electron distribution function.
The numeric value of the peak frequency
derived here, Eq. (10), is within 
a factor $\sim3$ of the values given in
\cite{AGSPN}, \cite{GPS} and \cite{AGWG}, and a factor $\sim10$ smaller than 
given in \cite{AGWb}. The discrepancy with \cite{AGWb} is due mainly to the 
fact that it is assumed in \cite{AGWb} that the spectral peak is close to
the synchrotron frequency of electrons with average Lorentz factor, while
the actual peak is closer to the synchrotron frequency of electrons
near the peak of the electron distribution function.

The break frequency (the frequency where the high-frequency spectrum changes the slope from $(p-1)/2$ to $p/2$ due to synchrotron cooling) is not prominent. The transition to the $p/2$ slope occurs in a manner that strongly depends on the details of the electron distribution function.

\subsection {Polarization}

If the observed afterglows are indeed synchrotron emission from ultrarelativistic blast waves propagating into ISM, the magnetic field needed to account for the emission must be generated by the blast wave. If the coherence length of the generated field grows at about the speed of light after the field was generated at the shock front, afterglows should be noticeably polarized.

\acknowledgements

We thank John Bahcall for a discussion that initiated this study. Our work 
was supported by NSF PHY-9513835. EW was also supported by the W. M. Keck 
Foundation.

\appendix 

\section{Ultrarelativistic blast wave.}
The Blandford \& McKee (1976) solution can be described as follows. 

Let $(t,r,\theta )$ be the space-time coordinates in the blast frame, $\theta$ 
is the polar angle which is assumed to be small, with $\theta =0$ in the 
observer direction. Let $E$ be the energy of the blast wave, and $n_i$ the 
unshocked ISM number density, $c=1$. The shock front propagates into the ISM 
with a Lorentz factor $\Gamma $ that decreases with time according to 
\begin{equation}
\Gamma ^2t^3={17\over 8\pi}{E\over n_im_p}\equiv T^3.
\end{equation}
Define a similarity variable 
\begin{equation}
\chi \equiv 8\Gamma ^2(1-{r\over t}).
\end{equation}
The shocked region is $\chi >1$, and the fluid flow in the shocked region is 
given by
\begin{equation}
\gamma ^2={1\over 2}\Gamma ^2\chi ^{-1},
\end{equation}
\begin{equation}
e=2\Gamma ^2\chi ^{-17/12}n_im_p,
\end{equation}
\begin{equation}
n=2\sqrt{2} \Gamma \chi ^{-5/4}n_i.
\end{equation}
Here $\gamma $ is the Lorentz factor of the flow, $e$ is the proper energy 
density, $n$ is the proper number density.

In Appendix B, we use $(\Gamma ,\gamma ,t_{\rm obs})$ as independent variables 
instead of $(t,r,\theta )$. Here $t_{\rm obs}$ is the time at which a photon 
emitted at $(t,r,\theta )$ is observed; with sufficient accuracy,
\begin{equation}
t_{\rm obs}=t-r+r{\theta ^2\over 2}=t-r+t{\theta ^2\over 2}.
\end{equation}
The coordinate transformation is (old coordinates in terms of new coordinates)
\begin{equation}
t=T\Gamma ^{-2/3},~~~~~~~~~~~~~~~~~~~~~~~~~~~~~~~
\end{equation}
\begin{equation}
r=(1-{1\over 16\gamma ^2})t=(1-{1\over 16\gamma ^2})T\Gamma ^{-2/3},
\end{equation}
\begin{equation}
{\theta ^2\over 2}={t_{\rm obs}\over t}-{1\over 16\gamma ^2}={t_{\rm obs}\over 
T}\Gamma ^{2/3}-{1\over 16\gamma ^2}.~
\end{equation}

From (A3)
\begin{equation}
\chi ={\Gamma ^2\over 2\gamma ^2}.
\end{equation}
With $\chi$ from (A10), expressions (A4), (A5) give the dependent quantities, 
energy density and density, in terms of new independent variables. In new 
coordinates, the space-time domain of the shocked fluid ($t>0$, $\chi >1$, 
$\theta >0$) is given by
\begin{equation}
\infty ~>~t_{\rm obs}~>~0,~~~~~~~~~~~~~~~~~~~~~
\end{equation}
\begin{equation}
\infty ~>~\Gamma~>~\left( {8t_{\rm obs}\over T}\right) ^{-3/8},~~~~~~~
\end{equation}
\begin{equation}
{\Gamma \over \sqrt{2} }~>~\gamma~>~{1\over 4}\left( {t_{\rm obs}\over 
T}\right) ^{-1/2}\Gamma ^{-1/3}.
\end{equation}

\section{Lightcurve}
Here we calculate the lightcurve of synchrotron emission from an 
ultrarelativistic blast wave. The physical assumptions of the model are 
discussed in the main text. 

It is convenient to start from the following expression for the total emitted 
energy 
\begin{equation}
E_r=\int r^2dr\int 2\pi \theta d\theta \int dt ~n\int d\omega 'P(\omega 
',\gamma _e,B){\omega \over \omega '} ~ {d\Omega '\over d\Omega }.
\end{equation}
The factors in (B1) are:
\begin{enumerate} 

\item Total number of emitting electrons at a given time 
\begin{equation}
\int r^2dr2\pi \theta d\theta ~\gamma n.
\end{equation}

\item Total energy emitted by one electron
\begin{equation}
\int {dt \over \gamma} \int d\omega 'P(\omega ',\gamma _e,B){\omega \over 
\omega '},
\end{equation}
where $\omega$ is the photon frequency in the burst frame, and $\omega '$ is 
the frequency in the local rest frame,
\begin{equation}
\omega '={1+\gamma ^2\theta ^2\over 2\gamma }\omega.
\end{equation}
$P(\omega ',\gamma _e,B)$ is the synchrotron radiation spectral power in the 
local rest frame emitted at the frequency $\omega '$, by one electron from a 
distribution with a mean Lorentz factor $\gamma _e$, in the magnetic field 
$B$. It is given by (Rybicki \& Lightman 1979)
\begin{equation}
P(\omega ',\gamma _e,B)={\sqrt{3}e^3\over 2\pi m_ec^2}BF({\omega '\over \omega 
_c}),
\end{equation}
\begin{equation}
\omega _c(\gamma _e,B)={3e\over 2m_ec}B\gamma _e^2,
\end{equation}
The dependence of emission on the mean Lorentz factor of electrons is shown 
explicitly. The dependence on the detailed distribution function of electrons 
is hidden in the definition of the synchrotron emission function $F(x)$. 
Namely, we define
\begin{equation}
F(x)=\int dzf_e(z)F_0({x\over z^2}),
\end{equation}
with $F_0(x)$ being the standard synchrotron emission function 
\begin{equation}
F_0(x)\equiv x\int_x^{\infty }d\xi ~K_{5/3}(\xi).
\end{equation}
In (B7), the normalized electron distribution function $f_e$ is defined by the 
following expression for the probability for the electron to have a Lorentz 
factor $\gamma _{el}$
\begin{equation}
{d{\rm Probability} \over d\gamma _{el}}={1\over \gamma _e}f_e\left( {\gamma 
_{el}\over \gamma _e}\right) .
\end{equation}

\item  The last factor in (B1) is the ratio of infinitesimal solid angles in 
the local rest and blast frames:
\begin{equation}
{d\Omega '\over d\Omega }={4\gamma ^2\over (1+\gamma ^2\theta ^2)^2}.
\end{equation}

\end{enumerate}

We assume that magnetic fields and electrons take up a fixed fraction of the 
proper energy density $e(r,t)$: 
\begin{equation}
{B^2\over 8\pi }=\xi _Be, 
\end{equation}
and 
\begin{equation}
\gamma _enm_ec^2=\xi _ee.
\end{equation}
We also assume that the normalized electron distribution function $f_e(z)$ in 
the shocked ISM is fixed. These assumptions might be approximately correct 
when synchrotron cooling becomes unimportant at later stages of the afterglow.

We use the Blandford \& McKee (1976) selfsimilar solution for the Lorentz 
factor $\gamma$, density $n$, and energy density $e$, and change the 
independent variables in (B1) from $(t,r,\theta ,\omega ')$ to $(\Gamma 
,\gamma ,t_{\rm obs}, \omega)$. Using Appendix A, we get 
\begin{eqnarray}
E_r={17\over 48}{E\over n_im_pc^2}\int_0^{\infty }d\omega \int_0^{\infty } 
dt_{\rm obs}~~~~~~~~~~~~~~~~~~~~~~~~~~~~~~~~~~~~~~~~~~~~~~~~~~~~~~~~~~~~~~
\nonumber \\~~~~~~~~~~~~~~~~~
\int_{(8t_{\rm obs}/T)^{-3/8}}^{\infty } d\Gamma ~\Gamma ^{-3}~\int_{(t_{\rm 
obs}/T)^{-1/2}\Gamma ^{-1/3}/4}^{\Gamma /\sqrt{2} } d\gamma ~\gamma 
^{-3}~nD^2P(D^{-1}\omega ,\gamma _e,B).
\end{eqnarray}
Here $D$ is the ``Doppler factor''
\begin{equation}
D={2\gamma \over 1+\gamma ^2\theta ^2}=2\gamma \left ({7\over 8}+2{t_{\rm 
obs}\over T}\Gamma ^{2/3}\gamma ^2\right) ^{-1},
\end{equation}
and $T$ is the characteristic time of the blast wave introduced in Appendix A. 

The spectral lightcurve is defined as luminosity per unit frequency:
\begin{equation}
L_{\omega }(t_{\rm obs })\equiv {dE_r\over dt_{\rm obs}d\omega }.
\end{equation}
From (A14),
\begin{equation}
L_{\omega }(t_{\rm obs })={17\over 48}{E\over n_im_pc^2}\int_{(8t_{\rm 
obs}/T)^{-3/8}}^{\infty } d\Gamma ~\Gamma ^{-3}~\int_{(t_{\rm 
obs}/T)^{-1/2}\Gamma ^{-1/3}/4}^{\Gamma /\sqrt{2} } d\gamma ~\gamma 
^{-3}~nD^2P(D^{-1}\omega ,\gamma _e,B).
\end{equation}
Now we use expression (B5) for the synchrotron power $P$, (B11) and (B12) for 
$B$ and $\gamma _e$, (A4), (A5), (A10) for $e$ and $n$.  Also, from now on we 
will denote by $t_o$ the observed time measured in units of $T$. We also 
define the frequency and spectral luminosity units, equations (4),(5). These 
are devised to get rid of constant factors in the resulting expression for the 
luminosity. We denote the frequency $\omega$ in units of $\omega _0$ by 
$\omega$, and the spectral luminosity $L_{\omega }(t_{\rm obs })$ in units of 
$E_0$ by $L_{\omega }(t_o)$. Then (B16) takes the following form
\begin{equation}
L_{\omega }(t_o)=\int_{(8t_o)^{-3/8}}^{\infty } d\Gamma ~\Gamma 
^{-1}\int_{t_o^{-1/2}\Gamma ^{-1/3}/4}^{\Gamma /\sqrt{2} } d\gamma ~\gamma 
^{-3}\left( {\Gamma ^2\over 2\gamma ^2}\right) ^{-47/24}D^2F[~D^{-1}\omega 
\Gamma ^{-3}\left( {\Gamma ^2\over 2\gamma ^2}\right) ^{25/24}~].
\end{equation}

Define new integration variables $x$ and $y$: $\gamma \equiv (\Gamma /\sqrt{2} 
)y$, $\Gamma \equiv (8t_o)^{-3/8}x^{-3/4}$. We obtain a selfsimilar spectral 
lightcurve 
\begin{equation}
L_{\omega }(t_o)=L_A(\omega t_o^{3/2}).
\end{equation}
Here
\begin{equation}
L_A(\omega )=192\int_0^1 dx ~x^{-1}~\int_x^1dy ~y^{35/12}(7+{y^2\over 
x^2})^{-2}F[~2x^3y^{-37/12}(7+{y^2\over x^2})\omega ~].
\end{equation}
With only an $\sim 2\% $ error in the resulting luminosity, we can replace the 
indices $35/12$ and $37/12$ by $3$, and get a simpler expression
\begin{equation}
L_A(\omega )=192\int_0^1dy~y^3\int_0^1 da ~a^3(1+7a^2)^{-2}F[~2a(1+7a^2)\omega 
~],
\end{equation}
where $a\equiv x/y$. From (B20), the adiabatic lightcurve is 
\begin{equation}
L_A(\omega )=48\int_0^1 da ~a^3(1+7a^2)^{-2}F[~2a(1+7a^2)\omega ~].
\end{equation}

\section{Synchrotron cooling}
Synchrotron plus adiabatic cooling of an electron with Lorentz factor $\gamma 
_{el}$ is described by
\begin{equation}
{d\gamma _{el}\over d\tau }={1\over 3}{\gamma _{el}\over n}{dn\over d\tau 
}-{4\over 3}\sigma _Tc{B^2\over 8\pi m_ec^2}\gamma _{el}^2,
\end{equation}
where $\tau $ is the proper time of the fluid element at the electron's 
location. We have $d\tau =dt/\gamma $. From (A5), $d\ln n=d\ln \Gamma 
-(5/4)d\ln \chi$. From (D8), $d\ln \chi =4d\ln t$. Then
\begin{equation}
{d\gamma _{el}\over dn}={1\over 3}{\gamma _{el}\over n}-{4\over 3}\sigma 
_Tc{B^2\over 8\pi m_ec^2}\gamma _{el}^2{1\over \gamma n}\left( {d\ln \Gamma 
\over dt}-{5\over t}\right) ^{-1}.
\end{equation}
Using (A1), (B11)
\begin{equation}
{d\gamma _{el}\over dn}={1\over 3}{\gamma _{el}\over n}+{8\over 39}{\sigma 
_T\over m_ec}\xi _B\gamma _{el}^2{te\over \gamma n}.
\end{equation}
To integrate, we need to express $t$, $e$, and $\gamma$ in terms of $n$. Let 
$\gamma _0$, $n_0$ and $e_0$ be Lorentz factor, proper density, and proper 
energy density at the shock passage time $t_0$. From (A1), (A3)-(A5),
\begin{equation}
\gamma _0={1\over \sqrt{2} }\Gamma _0,~~~~~~~~~
\end{equation}
\begin{equation}
n_0=2\sqrt{2} \Gamma _0n_i,~~~~~~~
\end{equation}
\begin{equation}
e_0=2\Gamma _0^2n_im_pc^2,~~~~
\end{equation}
where $\Gamma _0^2=T^3/t_0^3$. From (D8), (A1), (A3)-(A5),
\begin{equation}
\gamma =\gamma _0(t/t_0)^{-7/2},
\end{equation}
\begin{equation}
n=n_0(t/t_0)^{-13/2},
\end{equation}
\begin{equation}
e=e_0(t/t_0)^{-26/3}.
\end{equation}

Now (C3) can be written as 
\begin{equation}
{d\gamma _{el}\over dn}={1\over 3}{\gamma _{el}\over n}+{8\over 39}{\sigma 
_T\over m_ec}\xi _B{t_0e_0\over \gamma _0 n_0}\left( {n\over n_0}\right) 
^{-14/39}\gamma _{el}^2,
\end{equation}
and integrated
\begin{equation}
{1\over z}={1\over z_0}+{4\over 19}{\sigma _T\over m_e^2c^3}\xi _B\xi 
_e{t_0e_0^2\over \gamma _0 n_0}(1-y^{19/6}),
\end{equation}
Here $z\equiv \gamma _{el}/\gamma _e$, with $\gamma _e$ defined by (B12), $y$ 
is defined by (D1). Plug in (C4)-(C6), express $\Gamma _0$ in terms of $a$, 
$y$, and $t_o$. We get 
\begin{equation}
z^{-1}=z_0^{-1}+A(8t_o)^{-1/2}a^{-1}y^{-2}(1-y^{19/6}),
\end{equation}
$A$ is defined by (9), $a$ is defined by (D2). With the synchrotron cooling 
given by (C12), the spectral luminosity is
\begin{equation}
L_{\omega }(t)=192\int_0^1dy~y^3\int_0^1 da ~a^3(1+7a^2)^{-2}\int 
dz_0f_e(z_0)F_0[~2a(1+7a^2)\omega t_o^{3/2}/z^2~].
\end{equation}

\section{Transverse distances and proper times}
Adiabatic lightcurve (B20) is an integral over dimensionless variables $y$ and 
$a$:
\begin{equation}
y={\sqrt{2} \gamma \over \Gamma },~~~~~~~~
\end{equation}
\begin{equation}
ay=(8t_o)^{-1/2}\Gamma ^{-4/3}.
\end{equation}
To calculate polarization using (B20), we have to express the distance from 
the observer - burst center line $h$, and the proper time since the shock 
passage $\tau$, in terms of $y$ and $a$. This is done here.

The transverse distance is $h=r\theta$, and from (A8), (A9)
\begin{equation}
h=\sqrt{2} T\Gamma ^{-2/3}(t_o\Gamma ^{2/3}-{1\over 16\gamma ^2})^{1/2},
\end{equation}
and from (D1), (D2)
\begin{equation}
h={1\over 2}T(8t_o)^{5/8}(ay)^{1/4}(1-a^2)^{1/2}.
\end{equation}

Now the proper time... Equation of motion of a shocked particle is
\begin{equation}
{dr\over dt}=1-{1\over 2\gamma ^2}=8{r\over t}-7,
\end{equation}
here $t$ is the burst frame coordinate time. Integration gives $r=t-Ct^8$. 
Since the shock front is at
\begin{equation}
R=t-{t^4\over 8T^3},
\end{equation}
we get
\begin{equation}
r=t-{t^8\over 8T^3t_0^4},
\end{equation}
where $t_0$ is the burst coordinate time at which the particle was shocked. 
The similarity variable at the particle is 
\begin{equation}
\chi =t^4/t_0^4, 
\end{equation}
and
\begin{equation}
\tau=\int {dt\over \gamma}=\sqrt{2}\int dt{\chi ^{1/2}\over 
\Gamma}={2\sqrt{2}\over 9T^{3/2}t_0^2}(t^{9/2}-t_0^{9/2}).
\end{equation}
Using (D8), (A10), (A7), we get
\begin{equation}
\tau={2\sqrt{2}\over 9}T(8t_o)^{5/8}a^{5/4}y^{1/4}(1-y^{9/4}).
\end{equation}

\section{Degree of Polarization}
To estimate the degree of polarization $\Pi$ we use the adiabatic lightcurve 
and assume $F(\omega )\sim \omega ^{-s}$ in (B20). The latter simplification 
should not lead to a large error, because $\Pi$ turns out to be approximately 
the same in the relevant range $1>s>-1/3$. With these assumptions, the degree 
of polarization can be estimated as
\begin{equation} 
\Pi = {s+1\over s+5/3}{C_{LL}^{1/2}\over L}.
\end{equation}
Here the first factor is polarization of a power-law emission from one patch, 
$L$ is the total unpolarized luminosity, and $C_{LL}$ is the polarized 
luminosity correlator. Up to an irrelevant factor, 
\begin{equation}
L={1\over 4}\int daa^{3-s}(1+7a^2)^{-2-s},
\end{equation}
\begin{equation}
C_{LL}=\int da_1dy_1da_2dy_2(y_1y_2)^3(a_1a_2)^{3-s}[(1+7a_1^2)(1+7a_2^2)]^{-2-
s}~C_{12}~{\rm min}(1,{\epsilon _h\tau \over 2\pi h}).
\end{equation}
Here the $min$-term comes from the azimuthal angle integral, $\tau =(\tau 
_1+\tau _2)/2$, $h=(h_1+h_2)/2$. $C_{12}$ is the normalized magnetic field 
correlator, for which we take a simple form corresponding to equations 
(14),(15).
\begin{equation}
C_{12}=\theta (|\tau _1-\tau _2|-\epsilon _{\tau }\tau )\theta 
(|h_1-h_2|-\epsilon _h\tau ).
\end{equation}
We calculated (E1) numerically for different values of $s$, $\epsilon _h$ and 
$\epsilon _{\tau }$.

\begin{figure}[htb]
\psfig{figure=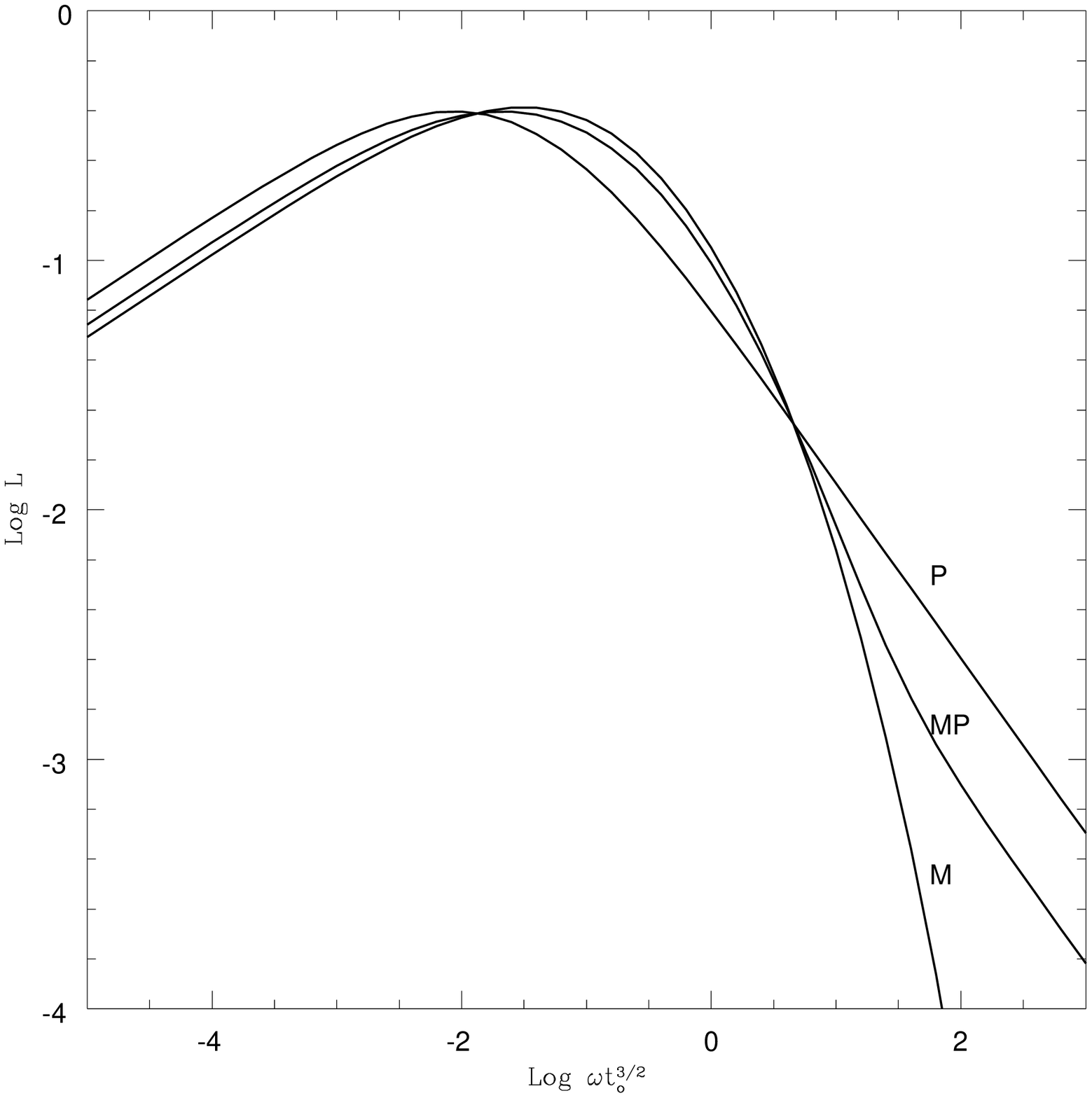,width=7in}
\caption{Adiabatic lightcurves (6) for different electron distribution 
functions (\S 3): power-law P, Maxwellian M, and mixed MP. This graph can be interpreted as luminosity at a given frequency as a function of time, or as the spectrum at a given time.}
\end{figure}

\begin{figure}[htb]
\psfig{figure=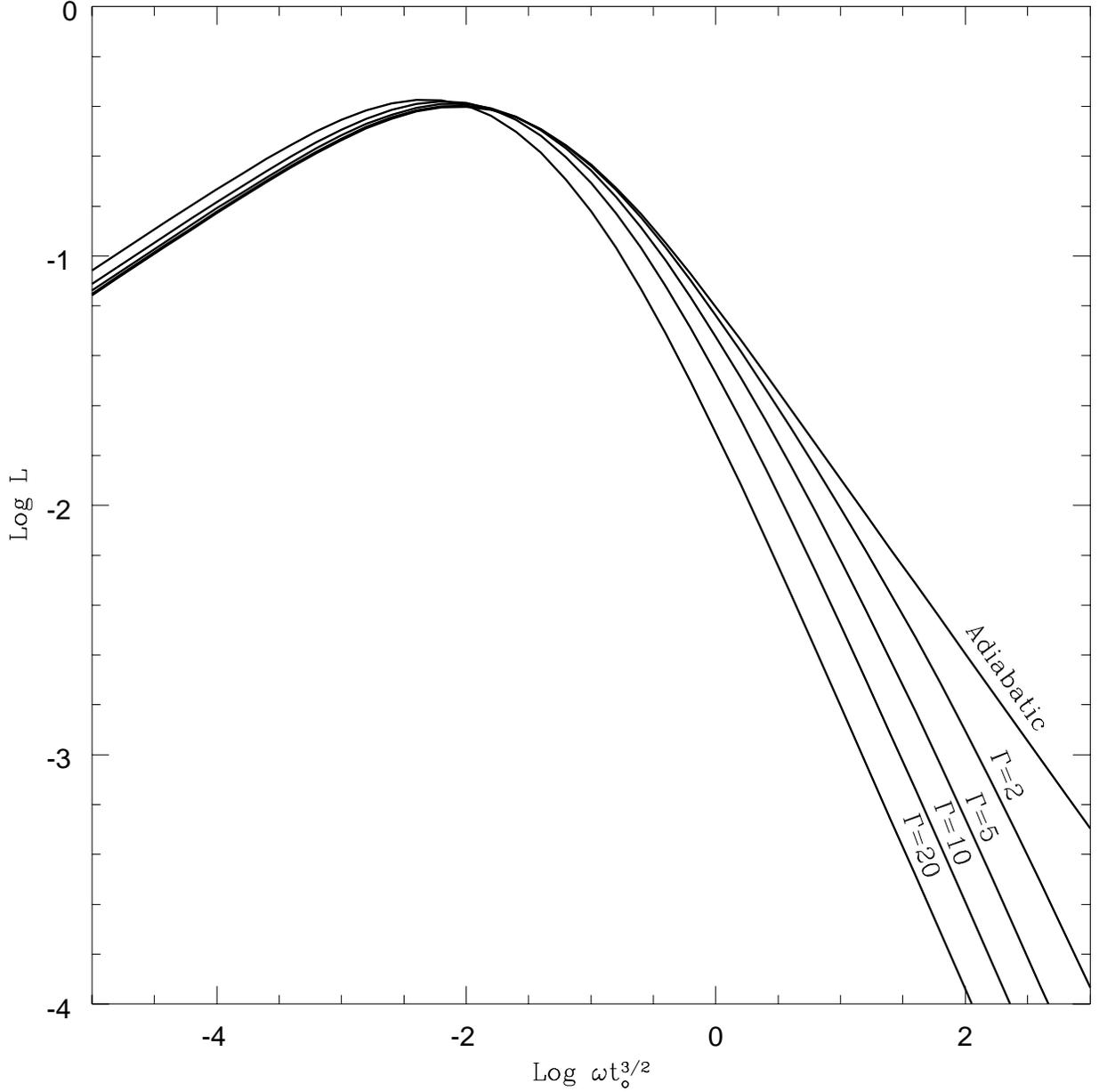,width=7in}
\caption{Nonadiabatic lightcurves (7) for $A=0.01$, for different observed 
times, and for a power-law electron distribution function (\S 3). Adiabatic 
lightcurve is shown for comparison. Nonadiabatic curves are marked by the 
Lorentz factors of the shock front at the time when observed photons were 
emitted from the shock front from $\theta =0$. Observed time in days is $t_{\rm day}=80 (E_{52}/n_1)^{1/3}\Gamma ^{-8/3}$.}
\end{figure}

\begin{figure}[htb]
\psfig{figure=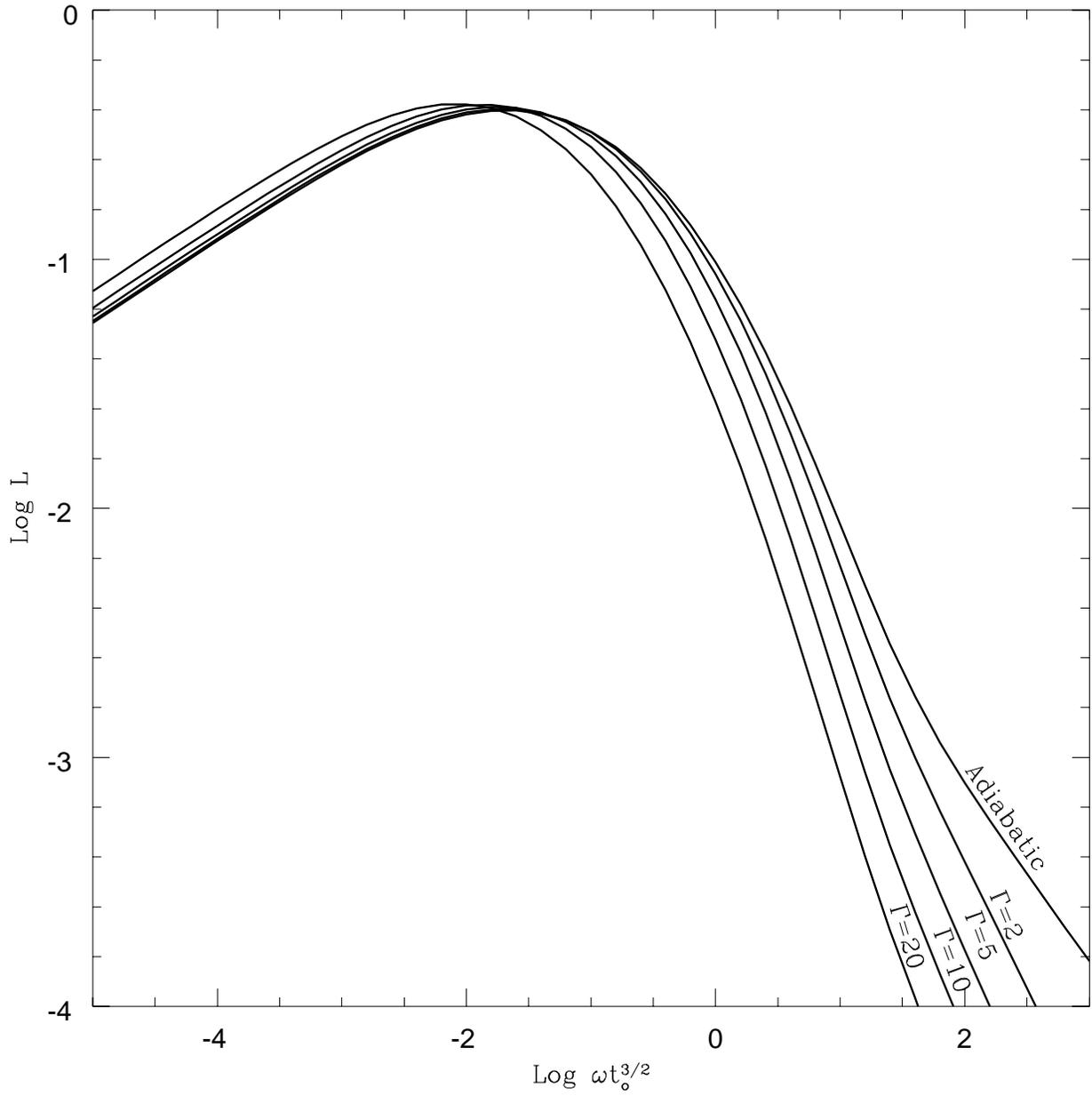,width=7in}
\caption{Same as Fig. 2, but for the mixed (power-law plus Maxwellian) electron distribution function (\S 3).}
\end{figure}

\end{document}